\documentclass[aps,prc,twocolumn]{revtex4}

\usepackage{graphicx}

\begin{document}


\title{The diffusive instability of kaon condensate in neutron star matter}

\author{Sebastian Kubis}
\email{sebastian.kubis@ifj.edu.pl}
\affiliation{Institute of Nuclear Physics Polish Academy of Sciences\\
ul.Radzikowskiego 152, 31-342 Krak\'ow, Poland\\}
\affiliation{Scuola Internazionale Superiore di Studi Avanzati\\ 
via Beirut 4, 34014 Trieste, Italy}
\date{\today}

\begin{abstract}
The beta equilibrated dense matter with kaon condensate is analysed
with respect to extended stability conditions including charge fluctuations.
This kind of the diffusive instability, appeared to be common property  in the
kaon condensation case. Results for three different nuclear models  are
presented.
\end{abstract}


\maketitle


\def\ep{\varepsilon}
\def\pa{\partial}
\def\th{\theta}
\def\chiv{\chi_{\raisebox{-.5ex}{\it\footnotesize v}}}
\def\chip{\chi_{\raisebox{-.8ex}{\it\footnotesize P}}}
\def\cp{c_{\raisebox{-.5ex}{\it\footnotesize P}}}
\def\kappat{\kappa_{\raisebox{-.5ex}{\it\footnotesize T}}}
\def\ni{\noindent}
\def\fm3{\rm ~fm^3}
\def\MeV{\rm ~MeV}

\newcommand{\beq}{\begin{equation}}
\newcommand{\eeq}{\end{equation}}
\newcommand{\beqa}{\begin{eqnarray}}
\newcommand{\eeqa}{\end{eqnarray}}
\newcommand{\arl}{\begin{array}{l}}
\newcommand{\earl}{\end{array}}
\newcommand{\fract}[2]{{\textstyle\frac{#1}{#2}}}
\newcommand{\fracd}[2]{{\displaystyle\frac{#1}{#2}}}
\newcommand{\Tr}{{\mathrm Tr}}
\newcommand{\oht}{\textstyle{\frac{1}{2}}}
\newcommand{\derc}[3]{\left( \frac{\partial #1}{\partial #2} \right)_{#3} }


\section{Introduction \label{introduction}}

Following the pioneering work of Kaplan and Nelson \cite{Kaplan:yq}, the
possibility of having kaon condensation in neutron stars has attracted a
lot of attention during past few years. We can divide the various proposed
models (for a review see \cite{Lee:ef}) into two groups: those coming from
effective chiral theory and those coming from the meson exchange model.
The former ones, based on chiral symmetry $SU(3)_L\times SU(3)_R$, treat
kaons as members of the Goldstone boson octet and couples them to baryons
according to the prescription appropriate for the nonlinear representation
of the chiral symmetry group \cite{CCWZ}. For the second group, kaons are
not coupled to baryons but only to heavy mesons \footnote{"Heavy mesons"
in the language of chiral symmetry means mesonic degrees of freedom which
are multiplets of the spontaneously unbroken flavour subgroup of the full
chiral group} ($\omega, \rho,\sigma$) and these in turn interact with the
baryons \cite{Mueller-Groeling:cw,Knorren:1995ds}. Although both
approaches give similar predictions for the onset of kaon condensation,
the matter composition remains uncertain in the chiral approach due to the
shape of the nuclear symmetry energy (an important quantity describing the
pure nucleon interaction in the isovector channel) not being
well-established.  Interesting results, among them the prediction of the
complete protonization of the stellar core, were presented in
\cite{Kubis:2002dr}.  However, there is also a {\em qualitative}
difference between the chiral and meson exchange models from a
thermodynamical point of view which we would like to stress in the present
paper.

Since kaon-nucleon interactions are strongly attractive, the presence of a
kaon condensate in dense matter makes the EOS softer. Because the K-N
coupling strength is not well known, usually a range of values for it are
taken into account. For a sufficiently strong, but still reasonable, K-N
coupling the pressure-energy relation, $p(\ep)$, exhibits a region where
the compressibility of the matter is negative. This is the signal of phase
separation and means that one needs to make a kind of "construction" which
ensures that the compressibility is positive or at least zero (as in the
Maxwell construction for the van der Walls equation of state). For neutron
star matter, the correct approach consists of applying the so-called Gibbs
construction. The system may separate into different phases provided that
their pressures and all relevant chemical potentials are equal (the Gibbs
conditions). The relevant chemical potentials are those connected with
quantum numbers which are conserved by the interactions maintaining the
equilibrium. In neutron star matter there are two such quantities: the
baryon number $B$ and the charge $Q$. This means that the corresponding
chemical potentials, for neutrons $\mu_n$ and for electrons $\mu_e$, must
be continuous across the phase boundary
\beq 
\mu^I_n = \mu^{II}_n ~~~~~~ \mu^I_e = \mu^{II}_e~.
\label{gibbs_mu}
\eeq
These equalities state that the phases are in {\em chemical} equilibrium
- there is no flow of baryon number or charge across the phase boundary.  
Similarly, pressure equality
\beq
P^I = P^{II} 
\label{gibbs_P}
\eeq  
states the condition for {\em mechanical} equilibrium - the phase
boundary stays at rest with respect to the matter. If the Gibbs conditions
(\ref{gibbs_mu},\ref{gibbs_P}) can be fulfilled, the system forms a
mixture of two phases with different densities of charge and baryon
number, occupying the volume in different proportions. This scenario, in
the kaon condensation case, was successfully presented in
\cite{Glendenning:1997ak}, where the meson exchange model was used. Mixed
phase formation has never been demonstrated for chiral models, although
the possibility of phase separation was noticed in \cite{Thorsson:bu}
where the Maxwell construction was used to remove the negative
compressibility region in the EOS. In the case of neutron star matter, the
Maxwell construction is usually made under the assumption that the matter
is locally neutral. This means that the construction ensures only the
equality of $P$ and $\mu_n$ whereas $\mu_e$ is different for the different
phases, so that the EOS obtained is not stable. The primary objective of
this work was to carry out a correct construction to enforce the Gibbs
conditions in the chiral model and get a stable EOS. However, a thorough
analysis shows that kaon condensation in the chiral approach leads to an
EOS which is completely unstable, independent of the strength of the
kaon-nucleon coupling. The same kind of stability considerations were also
applied to the meson exchange approach and we found that the stability of
an EOS with a kaon condensate depends not on the strength of $K$-$N$
interactions but on the manner in which they are realized.

This paper is organized as follows: in Sec.\ref{sec-stability}, the more
general stability conditions for beta stable matter are formulated. In
Sec.\ref{sec-chiral}, it is shown that it is not possible to make the
Gibbs construction for the chiral model and that the stability conditions
are not fulfilled. Then, in Sec.\ref{sec-meson-exch}, the same analysis is
repeated for two representative meson exchange models, one for which the
Gibbs construction had been performed earlier by other authors and one for
which that was not possible. In the final section, we discuss the results
obtained and the properties of Lagrangians which allow the instability to
be avoided.

\section{Intrinsic stability of a single phase under beta equilibrium
\label{sec-stability}}

The basic question in all neutron star matter research is the shape of the
relationship between pressure and energy density $p=p(\ep)$, usually
called the Equation of State. At zero temperature, the state of neutron
star matter should be uniquely described by quantities which are conserved
by processes leading to the equilibrium. For matter above nuclear density
$n_0 = 0.16 \fm3$ we have the nucleon beta cycles
\[
n \rightarrow p + l + \bar{\nu}_l ~~~~
p + l \rightarrow n + \nu_l ~~~{\rm where}~~l=e,~\mu
\]
which are governed by weak interactions. The baryon number $B$ is
conserved by this type of process so that the energy density $\ep$ and
pressure $P$ should be functions only of the baryon number density
$n_B=B/V$. However, this common statement includes the implicit assumption
that the matter is electrically neutral and spatially homogeneous. The
star as a whole must obviously be electrically neutral but the matter does
not need to be locally neutral.  There is no reason why the matter needs
to be homogeneous - it may separate into two or more phases with different
charge densities in such a way that the total volume on a large scale is
neutral. This scenario was presented in detail in
\cite{Glendenning:1992vb}. The second conserved quantity, electric charge,
should then be treated on the same basis as baryon number or, in other
words, the thermodynamic state of a given phase is described not by one
but {\em two} quantities: baryon number $B$ and charge $Q$, where $Q$ is
the sum over all charge carriers~\footnote{This quantity $Q$, which is
introduced with the opposite sign from usual, is positive for negative
charge carriers and so the chemical potential $\mu$ corresponding to $Q$
is just the electron chemical potential $\mu_e$ - a fundamental quantity
in beta equilibrium matter.\label{footq}} $Q=-N_p+N_e+N_\mu+\dots$
Starting from this point, we will formulate correct stability conditions
for a system which is not neutral, in general, but may have local charge
density different from 0. Consider a given phase with volume $V$: its
energy $U$ is the function of $V,B$ and $Q$, i.e. $U=U(V,B,Q)$. To
investigate the {\em intrinsic} stability of single phase it is more
convenient to introduce intensive rather than extensive quantities:
\[
u=U/B,~~~v=V/B,~~~q=Q/B
\]
and then, the energy per baryon becomes a function of two variables
\[
u=u(v,q)
\]
and the first law of thermodynamics takes the form
\beq
du = -P dv + \mu \;dq ~,
\label{1st-therm}
\eeq 
where the pressure and electrical chemical potential are:
\beq
P = - \derc{u}{v}{q} ~~,~~ \mu = \derc{u}{q}{v}
\label{P-mu}
\eeq
From the principle of minimum energy, it can be deduced that the phase is
intrinsically stable (i.e. it does not separate into different phases) if
and only if the energy per baryon is a {\em convex} function of its
variables $v$ and $q$ (for a detailed analysis, see \cite{callen}). In
terms of its differentials
\beq
d^2u = \oht u_{vv}\ dv^2 + u_{vq}\ dv\ dq + \oht u_{qq}\ dq^2 > 0.
\eeq
The sufficient condition for the quadratic form to be positive is that its
determinant must be positive and at least one of derivatives $u_{vv}$ or
$u_{qq}$ must be positive:
\beq
\left|\begin{array}{cc}
u_{vv}& u_{vq} \\
u_{vq}& u_{qq}
 \end{array} \right| > 0 ~~~{\rm and}~~~(~ u_{vv}>0 ~~{\rm or}~~ u_{qq}>0)
\eeq
Thermodynamical identities applied to the above second derivatives allow
us to express the positivity condition in terms of two equivalent pairs of
inequalities:
\beq
-\derc{P}{v}{q} > 0~~~~~~ \derc{\mu}{q}{P} > 0
\label{positivity1}
\eeq
\centerline{or}
\beq
- \derc{P}{v}{\mu} >0 ~~~~~~ \derc{\mu}{q}{v} > 0 
\label{positivity2}
\eeq
The pressure derivatives correspond to the well-known compressibilities of
matter, one with constant charge and the other with constant chemical
potential:
\beq
\kappa_q = -\frac{1}{v}\derc{v}{P}{q}~~~~
\kappa_\mu = -\frac{1}{v}\derc{v}{P}{\mu}
\eeq
The positivity of the compressibility, usually referred to as mechanical
stability, here means that baryon density fluctuations are stable.  The
physical content of the $\mu$ variation in (\ref{positivity1}) and
(\ref{positivity2}) becomes clear if we look carefully at their
dimensions. As $\mu$ has dimensions of energy/charge, these derivatives
have dimensions of $\rm energy/charge^2$ which is the inverse of the unit
of electrical capacitance. It is useful to introduce the {\em electrical
capacitance of matter}: $\chi$. Obviously, we must specify under which
constraints it is taken - under constant pressure or constant volume:
\beq
\chip = \derc{q}{\mu}{P}~~~~
\chiv = \derc{q}{\mu}{v}
\eeq
When $\chi$ is positive, it means the system is diffusively stable, i.e.
the charge fluctuations are stable. If $\chi$ changes its sign, the charge
fluctuations become infinite which means that the phase can be no longer
homogeneous and separates itself into parts with different charge density.
So, finally, it may be concluded that for stable matter, the
compressibility and electrical capacitance must be positive, what may be
expressed in two ways:
\beq
\left\{
\begin{array}{ccc}
\kappa_q &>& 0 \\ 
\chip &>& 0 
\end{array}
\right.~~~~~~{\rm or}~~~~~~
\left\{
\begin{array}{ccc}
\kappa_\mu &>& 0 \\ 
\chiv &>& 0 
\end{array}
\right.
\eeq
We can notice here some similarities to the one component system with
non-zero temperature, commonly treated in textbooks on thermodynamics. For
this kind of system, the internal energy $U$ is a function of volume,
entropy and particle number: $U(V,S,N)$. If we make the following
replacements:
\[
\begin{array}{ccc}
q & \rightarrow & s\\
\mu & \rightarrow & T\\
\end{array}
\]
i.e. the charge per baryon plays the role of specific entropy $s=S/N$ and
the chemical potential plays the role of temperature, then the electrical
capacitance $\chi_i$ corresponds to the specific heat:
\[
c_i = T \derc{s}{T}{i}~~~,~~i=v,P
\]
Instead of compressibility under fixed $q$ or $\mu$, we here have
compressibility under constant $T$ or $s$ and, in the same way, convexity
of the internal energy, which is now a function of specific volume and
entropy $u(v,s)$, requires the standard compressibility and specific heat
to be positive:
\[
\left\{
\begin{array}{ccc}
\kappat &>& 0 \\ 
c_v &>& 0 
\end{array}
\right.~~~~~~{\rm or}~~~~~~
\left\{
\begin{array}{ccc}
\kappa_s &>& 0 \\ 
\cp &>& 0 
\end{array}
\right.
\]
which are well known sufficient conditions for stability of a one
component system at non-zero temperature.

In the next section we stress that the positivity requirements on the
electric capacitance of matter may not always be ensured. This property
is strictly connected with the interactions used for the description of
the dense nuclear matter.

\section{The chiral model}
\label{sec-chiral}

To investigate phase stability in the chiral approach, we focus on the
model used in \cite{Thorsson:bu} where kaon-nucleon interactions are
described by the Lagrangian coming from chiral effective field theory
\beqa
{\mathcal  L_\chi} & = & \fract{f^2}{4}\Tr \pa_\mu U \pa^\mu U^\dagger +
        \Tr\bar{B}(i \gamma^\mu D_\mu -  m_B) B  \label{lan-KN} \\
   & &   +  F \Tr \bar{B}\gamma^\mu \gamma_5 [ {\mathcal  A}_\mu,B]
  +  D \Tr \bar{B}\gamma^\mu \gamma_5 \{ {\mathcal  A}_\mu,B\}
\nonumber \\
   & & + c\, \Tr {\mathcal   M} (U + U^\dagger) +
   a_1 \Tr \bar{B}(\xi {\mathcal  M} \xi + \xi^\dagger {\mathcal  M}
\xi^\dagger) B
\nonumber \\
   & & + a_2 \Tr \bar{B}B(\xi {\mathcal  M} \xi + \xi^\dagger {\mathcal
 M} \xi^\dagger) 
\nonumber \\
& &+  a_3 \Tr \bar{B}B\; \Tr(\xi {\mathcal  M} \xi + \xi^\dagger {\mathcal
M} \xi^\dagger)
.\nonumber
\eeqa
The only parameter in $\mathcal  L_\chi$ which is not well determined is 
$a_3$, which is connected with the kaon-nucleon sigma term 
$\Sigma_{KN} = \frac{1}{2} m_s \langle N | \bar{u} u + \bar{s}s | N 
\rangle = -(\fract{1}{2}a_1 + a_2 + 2 a_3) m_s $.
Reasonable values for $a_3 m_s$ are in the range $-310 \ \mathrm{MeV} <
a_3 m_s < -134 \ \mathrm{MeV}$ which corresponds to a strangeness content
for the proton within the range \mbox{$0.2 > {\langle \bar{s}s \rangle_p}
/ { \langle \bar{u}u + \bar{d}d + \bar{s}s\rangle_p} > 0 $}
\cite{Donoghue:1985bu}. The value of this parameter does not appear to be
of great importance for the following analysis of matter stability. The
energy density coming from ${\mathcal L_\chi}$ is found by means of Baym's
theorem, which states that $\langle K^- \rangle = \frac{f \th}{\sqrt{2}}
\exp(-\mu_K t)$ (assuming only the s-wave kaon-nucleon interaction):
\beq
\ep_{KN} = f^2 {\mu_K^2\over2} \sin^2 \th +
     2(m^2_K f^2 - n \Sigma_{KN}(x)) \sin^2{\th\over2}
\eeq
where
\[
\Sigma_{KN}(x) = -(a_1 x + a_2 + 2 a_3) m_s .
\]
The kaon-nucleon contribution must be properly minimized with respect to
$\th$, and this leads to the following equation for $\th$
\beq
\cos\th = \frac{1}{f^2 \mu_K^2} \left(m_K^2 f^2 - n \Sigma_{KN}(x)
    - \fract{1}{2}\mu_K n (1\!+\!x) \right).
\label{th-min}
\eeq
For the interactions in the pure nucleonic sector we adopt the
phenomenological formula \cite{Lattimer61} for the energy density
\beq \ep_{NN} = \frac{3}{5}
E_F^{(0)} n_0 (\fract{n}{n_0})^{5/3} + V(n) + m n +  n(1 - 2x)^2
E_s(n), \label{endens-NN}
\eeq
 where
\beqa
E_s(n) &=& \frac{3}{5}(2^{2/3} - 1) 
E_F^{(0)} ((\fract{n}{n_0})^{2/3}  - F(\fract{n}{n_0})) \nonumber \\ 
&& + E_s(n_0) F(\fract{n}{n_0}).
\label{ensymm}
\eeqa
There are two potential contributions in (\ref{endens-NN}) coming from the
isoscalar and isovector sectors of the nucleon interaction, $V(n)$ and
$E_s(n)$ respectively. $V(n)$ is parametrized to get correct values for
the compression modulus at the saturation point $K_0=240 \MeV$ and it mainly
determines the stiffness of the EOS. The nuclear symmetry energy $E_s(n)$
at the saturation point takes the value $30 \MeV$. The potential
contribution $F(u)$ is an unknown function which determines the behaviour
of $E_s$ at higher densities.  The importance of the symmetry energy for
the chemical composition of dense matter in the kaon condensate case was
presented in \cite{Kubis:2002dr}. Here we put $F(u)=u$ to mimic typical
results from relativistic mean field theory (RMF) since we are going to
compare the chiral model to the RMF approach only as regards the
difference in the kaon-nucleon interaction. The total energy density is a
sum
\beq
\ep = \ep_{KN} + \ep_{NN} + \ep_{lep},
\eeq
where the leptonic contribution is expressed in the standard way by means
of chemical potentials for electrons and muons
\beq
\ep_{lep} = \frac{\mu_e^4}{4 \pi^2} + m_\mu^4
g(\sqrt{\eta(\mu_\mu^2-m_\mu^2)}/m_\mu)~,
\eeq
where the function $g(t)$ is:
\beqa
g(t) & = & \frac{1}{8 \pi^2} ((2 t^3 + t) \sqrt{1 + t^2} -
\mathrm{arsinh} t )~,
\\
& & \eta(x) = \left\{ \arl x ~;~ x \geq 0 \\  0 ~;~ x \leq 0~~~. \earl
\right.
\nonumber
\eeqa
From the total energy of the system we are able to find all of the
necessary thermodynamic quantities such as the nucleon chemical potentials
and the pressure
\beqa
\mu_n\!&\!=\!&\! E_F^{(0)}(\fract{n}{n_0})^{2/3} + m + V'(n) +
n(1\!-\!2x)^2 E_s'(n) \label{mun} \\ 
&&+ (1\!-\!4x^2)E_s(n) - (2 \Sigma_{Kn}
+ \mu_K) \sin^2{\th\over 2}, \nonumber \\  
\mu_p\!&\!=\!&\!
E_F^{(0)}(\fract{n}{n_0})^{2/3} + m + V'(n)  + n(1\!-\!2x)^2 E_s'(n) 
 \\
&& -(1\!-\!2x)(3\!-\!2x)E_s(n) - (2 \Sigma_{Kp} + 2\mu_K)
\sin^2{\th\over 2},\nonumber\label{mup} \\
P\!&\!=\!&\! - \ep + \sum_i \mu_i n_i\nonumber = \\ 
& = & \frac{2}{5} E_F^{(0)} n_0 (\fract{n}{n_0})^{5/3} -
V(n) \nonumber \\ 
&& +\; n V'(n) +  n^2 (1-2 x)^2 E_s'(n) \nonumber \\
&& +\; \frac{1}{2} f^2 \mu_K^2
\sin^2\th - 2 m_K^2 f^2 \sin^2\frac{\th}{2} \label{P} \\
&& +\;   \frac{\mu_e^4}{12 \pi^2}
+   m_\mu^4 g_p\left(\sqrt{\eta(\mu_\mu^2-m_\mu^2)} /m_\mu\right)~~, \nonumber
\eeqa
where for the chemical potentials it is useful to introduce kaon-neutron
and kaon-proton sigma terms: $\Sigma_{Kn}=\Sigma_{KN}(0) ,
\Sigma_{Kp}=\Sigma_{KN}(1)$.

Now, the total energy and pressure are functions of five variables (the
number densities $n_n, n_p, \mu_K, \mu_e, \mu_\mu$ corresponding to the
five species of particles $p,n,K^-,e,\mu$ present in the system). By the
use of the beta equilibrium conditions
\beq
\mu_K = \mu_e = \mu_\mu =  \mu_n - \mu_p \label{beta-seq} 
\eeq
we may reduce these to two independent variables as mentioned in section 
\ref{sec-stability}. The above equations also allow us to show that
\[
\mu_e =  \derc{u}{q}{v}
\]
which justifies identifying $\mu_e$ with the charge chemical potential 
appearing in (\ref{P-mu})
\[
\mu \equiv \mu_e .
\]
Which two of the five variables are chosen is a matter of convenience. For
instance, in order to make the Gibbs construction we need the pressure as
function of the chemical potentials $P(\mu_n,\mu)$ for finding a state
preserving the Gibbs conditions (\ref{gibbs_mu},\ref{gibbs_P}) for the
two-phase coexistence. Usually, the compressibility of the neutral matter
is first tested $\derc{P}{v}{q=0}>0$, and following this procedure, for
the full stability considerations we have chosen the pair of inequalities
expressed by (\ref{positivity1}). For the $\mu$ derivative in
(\ref{positivity1}) it is useful to express the pressure as a function of
$\mu$ and $q$, where
\beq
q = \frac{-n x +f^2 \mu \sin^2\th + n (1+x)\sin^2\frac{\th}{2}+n_{lep}}{n}
\label{qperB}
\eeq
and then to look at the pressure contours in the $q$-$\mu$ plot. Using 
the condition of
beta equilibrium (\ref{beta-seq}), the 
pressure in Eq.(\ref{P}) can be written as a function of $n,x,\th,\mu$. 
These
quantities were found numerically for given $P$ and $q$ by solving Eqs 
(\ref{th-min},\ref{P},\ref{qperB}), with the proton fraction under beta
equilibrium being
\beq
x=\frac{1}{2} - \frac{\mu(1+\cos\th) + 2 a_1m_s(1-\cos\th)}{16 E_s(n)}.
\eeq

\section{Phase stability in chiral model}

\begin{figure*}[t]
\includegraphics[width=.59\textwidth]{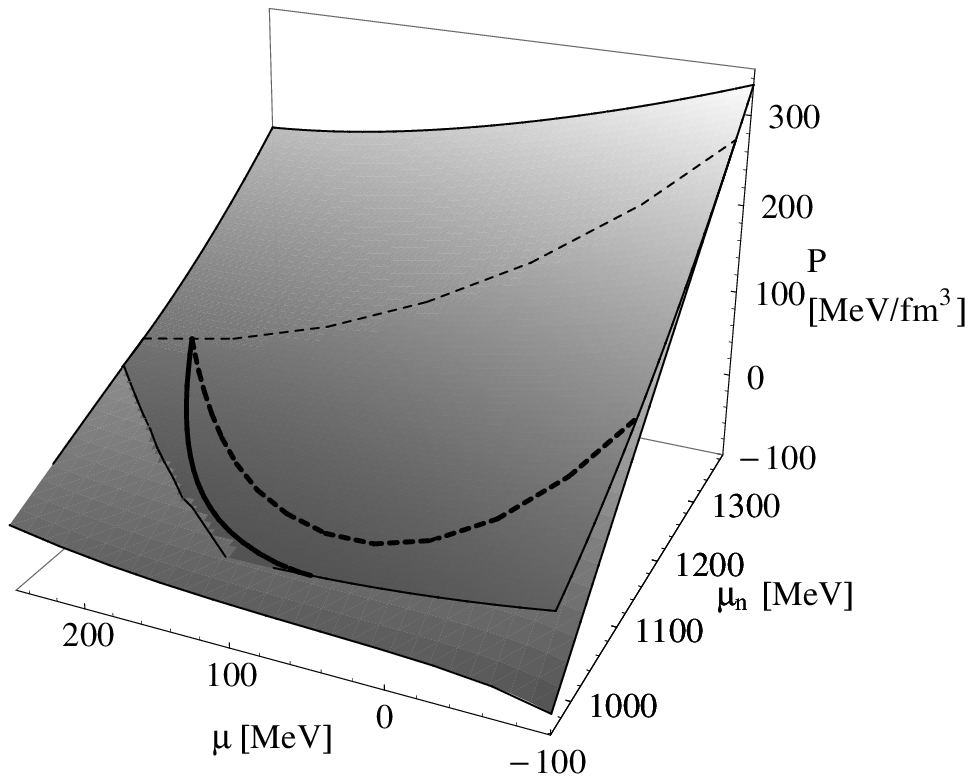}
\includegraphics[width=.4\textwidth]{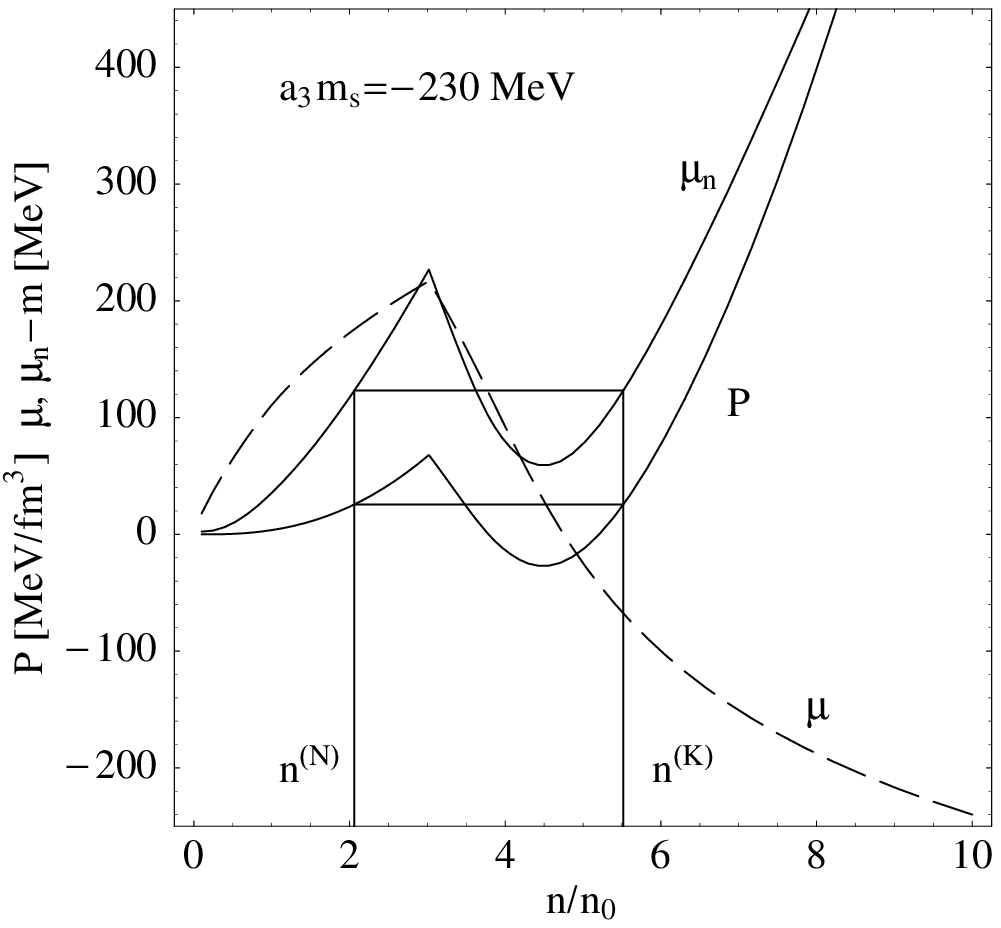}
\caption{Left panel: the pressure sheets for the normal phase (upper) and
the kaon phase (lower) for $a_3 m_s = -230$ MeV. The thin dashed line
corresponds to $\th=0$ where condensation starts and the two sheets are
tangent. The thick line represents the neutral matter where its dashed
part is the projection of the line which, in fact, lies on the lower sheet
corresponding to the kaon phase. The right-hand panel shows the Maxwell
construction which preserves the continuity only of $P$ and $\mu_n$.}
\label{fig-gchiral}
\end{figure*}

\begin{figure}[h]
\includegraphics[width=.45\textwidth]{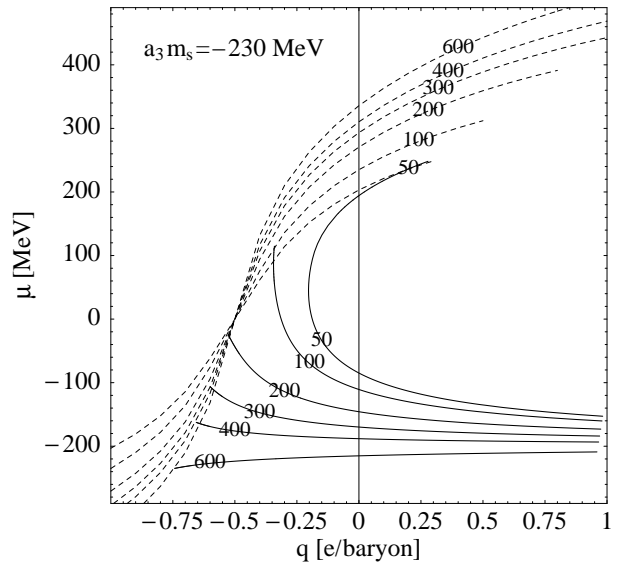}
\caption{Isobars on the $q$-$\mu$ plot for the normal phase (dashed lines) and 
the kaon phase (solid lines). 
The values of pressure indicated are expressed in $\rm MeV/fm^{3}$.}
\label{fig-isob_ch}
\end{figure}

\begin{figure*}[!]
\includegraphics[width=.43\textwidth]{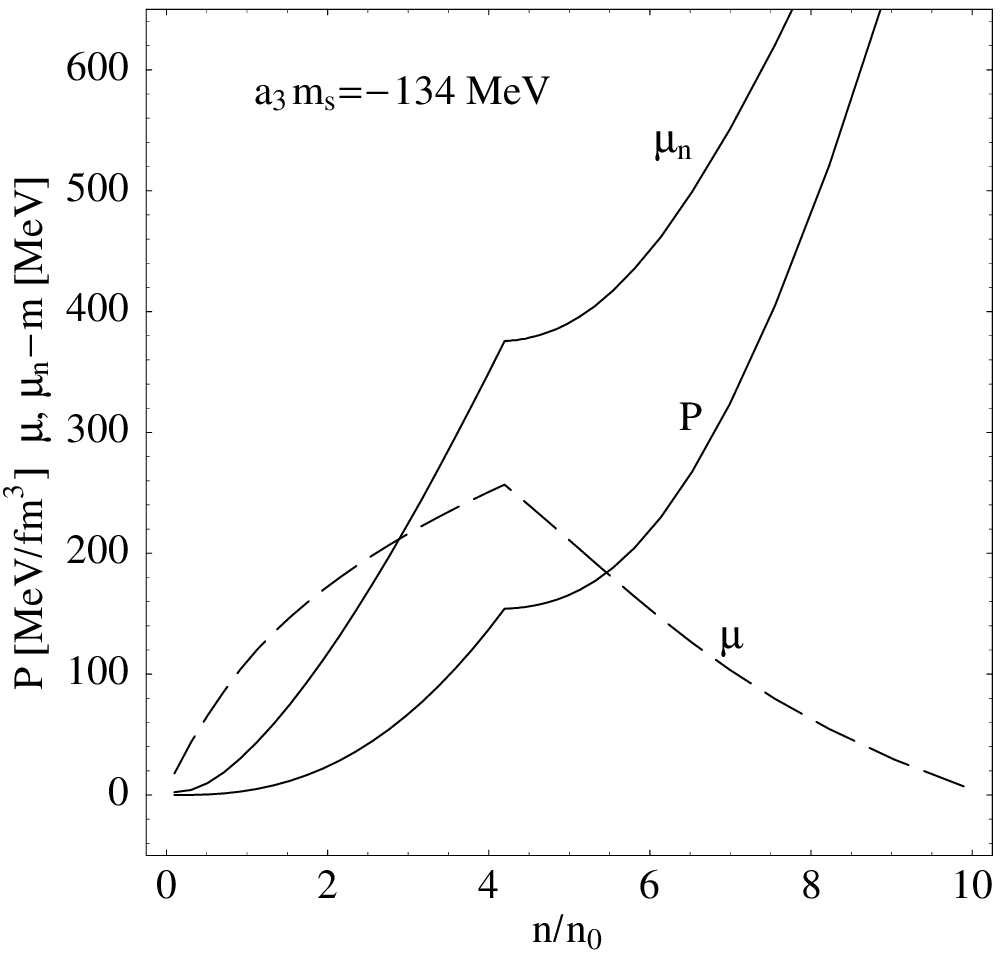}$~~$
\includegraphics[width=.43\textwidth]{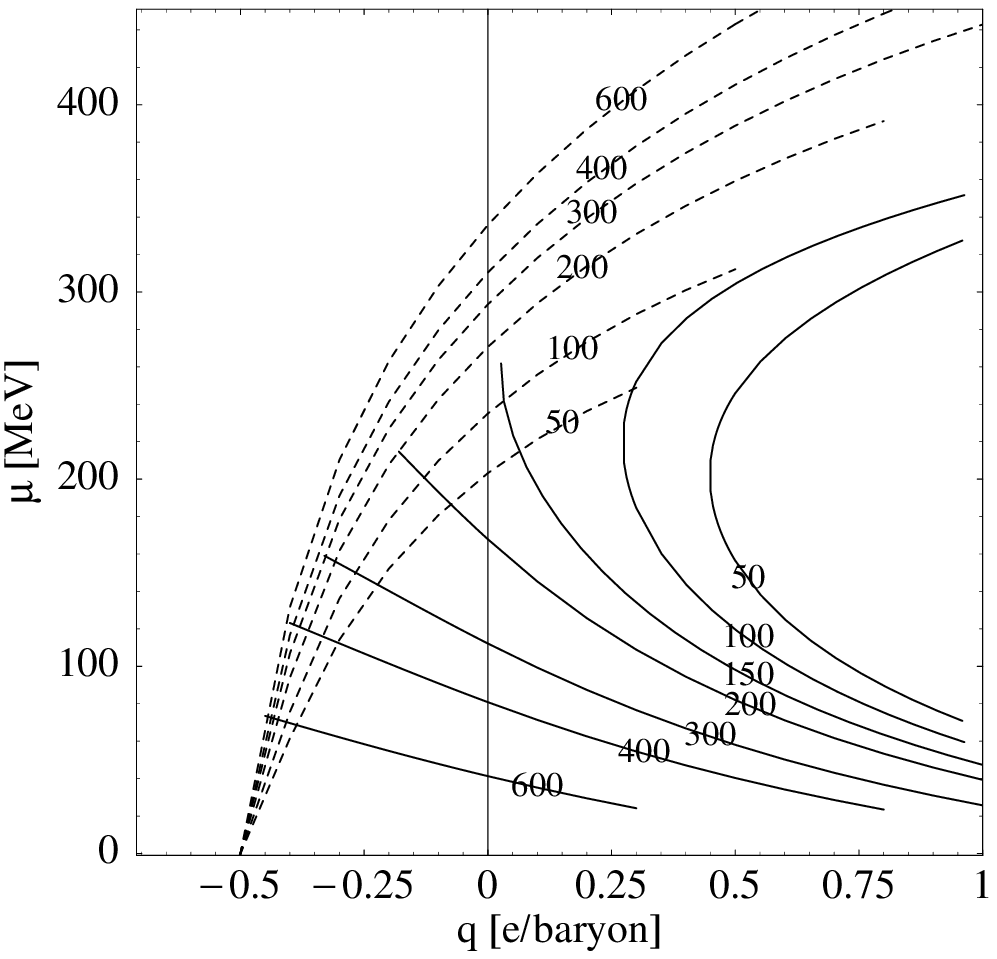}
\caption{Kaon phase instability for minimal $K$-$N$ coupling.}
\label{fig-isob_ch_min}
\end{figure*}

For the kaon condensation presented in \cite{Thorsson:bu}, the
pressure-density relation revealed a region of negative compressibility
when the matter was locally neutral \mbox{$q=0$}. This unstable region was
removed by means of the Maxwell construction which, as mentioned in the
introduction, does not preserve continuity of the charge chemical
potential. By abandoning the local neutrality of matter one may to try to
perform the Gibbs construction as presented in Fig.\ref{fig-gchiral} where
the pressure is shown as a function of the baryon density and the charge
chemical potential and also $\mu$ discontinuity is shown in the case of
the Maxwell construction.
From this figure one can see that the pressure of the kaon phase is always
lower than that of the normal phase. The only points where the two
pressures are equal is on the $\th=0$ line, where the condensate starts to
grow. However, at this line these two surfaces are tangent, or the
pressure gradients for both phases are equal:
\beq
\derc{P^{N}}{\mu}{\mu_n} = \derc{P^{K}}{\mu}{\mu_n}
\eeq
These pressure gradients are just the charge densities, according to the 
thermodynamical relation
\beq
dP = n\; d\mu_n + n_q\; d\mu
\eeq
where $n_q=Q/V$, so there is no difference in charge density between the
normal and kaon phases - the transition from the normal phase to the kaon
phase is of second order. The lack of charge difference between the phases
makes it impossible to have formation of the mixed phase along $\th=0$.
There is no other region in the ($\mu_n,\mu$) plane where the pressures
for the two phases might be equal. This is a signal of inconsistency
between pressure and energy for the kaon phase.  Thermodynamic principles
state that the system prefers the phase having greater pressure for given
chemical potentials ("pressure maximization") and simultaneously smaller
energy for given particle numbers ("energy minimization").  Going along
the neutrality line with increasing density, after the unstable region is
passed, the line should have entered into region where the mixed phase
disappears and the pure kaon phase is again preferable. Although for all
densities the kaon phase has lower energy, suggesting that it is the
preferred phase, its pressure $P^{K}$ never exceeds $P^{N}$. The
fundamental reason for this is that the kaon phase does not represent a
stable system at all. This conclusion is confirmed by testing the second
derivative of the energy, $\chip$, which is shown by means of constant
pressure contours in the $q$-$\mu$ plot, in Fig.\ref{fig-isob_ch}. The
isobars are represented keeping in mind the physical constraints:  $0<
x<1$ and $-1<cos\th<1$, and so usually they do not cover the whole area of
the plot.  As one can see, the isobar slope for the kaon phase is negative
for the neutral phase \mbox{$q=0$}. Moreover, a negative $\chip$ is also observed
for almost all values of $q$. For comparison, the isobars for the normal
phase are also plotted and they always exhibit a positive value of $\chip$
which means that the normal phase is always stable.
Such behaviour is not an effect of the kaon-nucleon coupling being too
strong. A negative $\chip$ is also observed for the minimal kaon-nucleon
coupling corresponding to the vanishing strangeness content of proton
where $a_3 m_s = -134 $ MeV.  The plots for this minimal coupling are
presented in Fig.\ref{fig-isob_ch_min}. The pressure-density relation for
the neutral matter does not reveal any region of negative compressibility,
as seen in the left panel in Fig.\ref{fig-isob_ch_min}. Nevertheless the
electrical capacity $\chip$ is still negative for \mbox{$q=0$}. This means that
the matter, although being mechanically stable, does not fulfil the
diffusive stability condition (\ref{positivity1}). Also, for most values
of $q$, kaonic matter is diffusively unstable except in a small region for
low values of pressure and only positive values of $q$. At this point we
conclude that, within the framework of the chiral model, matter with a
kaon condensate is intrinsically unstable with respect to charge
fluctuations for any values of $K$-$N$ coupling.

\section{Stability in the meson exchange model}
\label{sec-meson-exch}

Having discussed the stability considerations for the chiral model we now
want to address the same question for the second class of models used for
treating kaon condensation - the meson exchange models. The common feature
of this class of models is that kaons are coupled to vector or scalar
mesons but not directly to baryons. The way of making this coupling is not
unique. Let us focus on two different ways of coupling presented by
Knorren, Prakash and Ellis in \cite{Knorren:1995ds} and by Glendenning and
Schaffner-Bielich in \cite{Glendenning:1997ak}. In the following, we refer
to them as the KPE and GS approaches. For the GS model, the Gibbs
construction was successfully carried out whereas for KPE it was not
possible to do this as was later mentioned in \cite{Pons:2000iy}. The two
approaches differ regarding the pure nucleon interactions as well as the
kaon-nucleon interactions. In order to underline the role played by the
difference in the kaon-nucleon sector we have formulated two models which
are identical in their nucleon part but different in the kaon part. Hence,
for both of them we keep the same Lagrangian for the non-strange sector
\begin{eqnarray}
{\cal L}_N\!&\!=\!&\!
\bar N\!
\left(i \gamma_\mu \partial^\mu\! - m_N\!  + g_{\sigma N} \sigma\! - g_{\omega N}
\gamma_\mu V_\mu\! - g_{\rho N} \vec{\tau}_N \vec{R}_\mu \right)\! N \nonumber\\
&& +\frac{1}{2} \partial_\mu \sigma \partial^\mu \sigma  - \frac{1}{2} m_\sigma^2
\sigma^2 - U(\sigma) \nonumber \\
&& - \frac{1}{4} V_{\mu\nu} V^{\mu\nu}  + \frac{1}{2}
m_\omega^2 V_\mu V^\mu  \nonumber \\
&& - \frac{1}{4} \vec{R}_{\mu\nu} \vec{R}^{\mu\nu}  +
\frac{1}{2} m_\rho^2 \vec{R}_\mu \vec{R}^\mu , \label{lanN}
\end{eqnarray}
where $N$ denotes nucleons and $\sigma, V, \vec{R}$ are the scalar, vector 
and isovector meson fields (for details see \cite{Glendenning:1997ak}). In 
the kaonic sector we used two different Lagrangians:
\beqa
{\cal L}_K^{GS}~~ &=& {\cal D}_\mu^* \bar{K} {\cal D}^\mu K - {m^*_K}^2 \bar{K}K
\label{lanK-GS} \\
&&\nonumber\\
{\cal L}_K^{KPE} &=&
{\pa}_\mu K^+ {\pa}^\mu K^-  - {m^{*2}_K} K^+ K^-  \label{lanK-KPE}\\
&& + i(g_{\omega K} V^\mu + g_{\rho} R_3^\mu)(K^+
\pa_\mu K^- - K^- \pa_\mu K^+) \nonumber
\eeqa
where the covariant derivative is 
\beq 
{\cal D}_\mu =
\pa_\mu + i g_{\omega K} V_\mu +  i g_{\rho K} \vec{\tau}_K \vec{R}_\mu
\label{cov-der}
\eeq
The two Lagrangians have different mass terms:
\beq 
m^{*(GS)}_{K} = m_K - g_{\sigma K} \sigma 
\eeq
\beq
m^{*(KPE)}_{K} = \sqrt{m_K^2 - g_{\sigma K} m_K \sigma }
\eeq
Apart from the usual Weinberg-Tomozawa term (the second term in ${\cal
L}_K^{KPE}$), the covariant derivative introduces additional terms which
are quadratic in the meson fields $V^\mu$ and $\vec{R}^\mu$. These
additional terms modify the equation of motion for the vector mesons in
such a way that it couples to the conserved current
\beq
\pa_\nu V^{\mu\nu} + m_\omega V^\mu = g_{\omega N}
\bar{N}\gamma^\mu N -  g_{\omega K} J_K^\mu 
\eeq 
where the kaon current $J_K^\mu$, coming from the symmetry $K^\pm 
\rightarrow \exp^{\pm i \alpha} K^{\pm}$, is 
\beqa
J^K_\mu &=& iK^- \pa_\mu K^+ -i\pa_\mu K^-K^+ \nonumber \\
&& - 2g_{\omega K} V_\mu K^- K^+ - 2g_{\rho K} \vec{\tau}_K \vec{R}_\mu K^- K^+
\eeqa
In this scheme, the divergence of the vector field vanishes and so it is
possible to fulfil the Ward identity for a vector field in a medium, as
pointed out by Schaffner and Mishustin in \cite{Schaffner:1995th}. In the
KPE Lagrangian, the coupling to the scalar $\sigma$ is only linear whereas
for the GS Lagrangian a quadratic term in $\sigma$ is also present.

For both models presented here, we use the same parametrization for the
nucleon part (\ref{lanN}) which gives the properties of the saturation
point as follows: the binding energy at $n_0=0.153~{\rm fm^{-3}}$ is
$E/A=-16.3$ MeV, the symmetry energy $E_s(n_0) = 32.5$ MeV, the
compressibility $K_0=240$ MeV, and the nucleon effective mass
$m^*/m=0.78$. As shown in \cite{Kubis:1997ew}, the density dependence of
the symmetry energy $E_s$ in the RMF models is dominated by a linear
increase with density which is in agreement with the $E_s(n)$ introduced
in the chiral approach (\ref{ensymm}). In order to keep the same strength
for the $K$-$N$ interactions in both models, one may compare the optical
potential in symmetric nuclear matter ($R_{0,3}=0$) coming from
(\ref{lanK-KPE}) and (\ref{lanK-GS})
\beqa
U^{GS}_K~~ &=& - g_{\sigma K} m_\sigma - g_{\omega K} V_0 +
 \frac{(g_{\omega K}V_0)^2+(g_{\sigma K} \sigma)^2}{2 m_K}\nonumber\\
&& \\
U^{KPE}_K &=& - \frac{1}{2} g_{\sigma K} m_\sigma - g_{\omega K} V_0
\eeqa
At nuclear density $n_0$, where $U_K$ is fixed, the quadratic terms in
$U^{GS}$ contribute no more than a few percent to the total and so we
neglect them, as was also done in \cite{Glendenning:1997ak}. The only
difference is then in the scalar coupling constant $g_{\sigma K}$ which is
twice as large in the KPE parametrization as in the GS one. The value of
the optical potential may also be used to relate the $K$-$N$ coupling
strength to the chiral model for which $U_K$ is
\beq 
U_K^{\chi} = -\frac{\Sigma_{KN}(\frac{1}{2})\, n_s}{2 m_K f^2_\pi} -
\frac{3n}{8f^2_\pi} ~.
\eeq
A strangeness content of the proton within the range 0-0.2 then
corresponds to $U_K$ being between -73 and -120 MeV. To keep contact with
the results of Glendenning and Schaffner, we use the values -80 and -120
MeV in our further analysis.

\begin{figure*}[!]
\includegraphics[width=.43\textwidth]{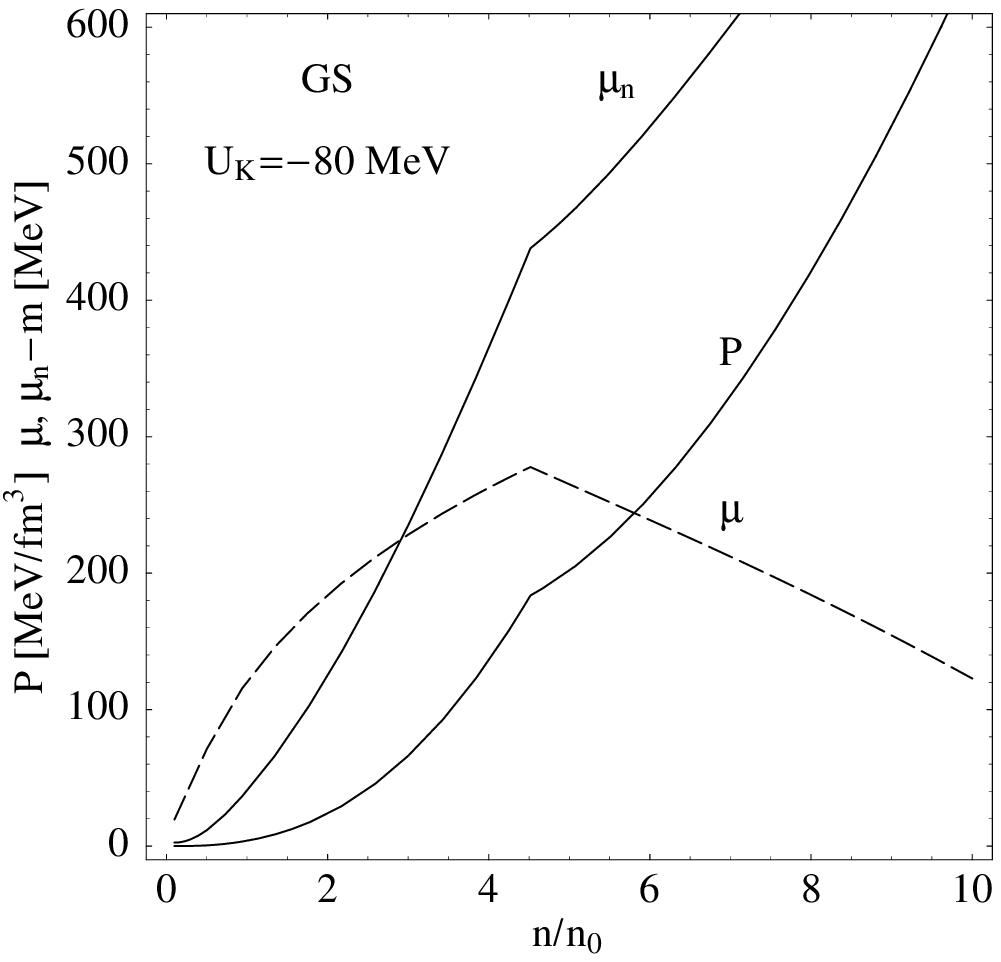}$~$
\includegraphics[width=.422\textwidth]{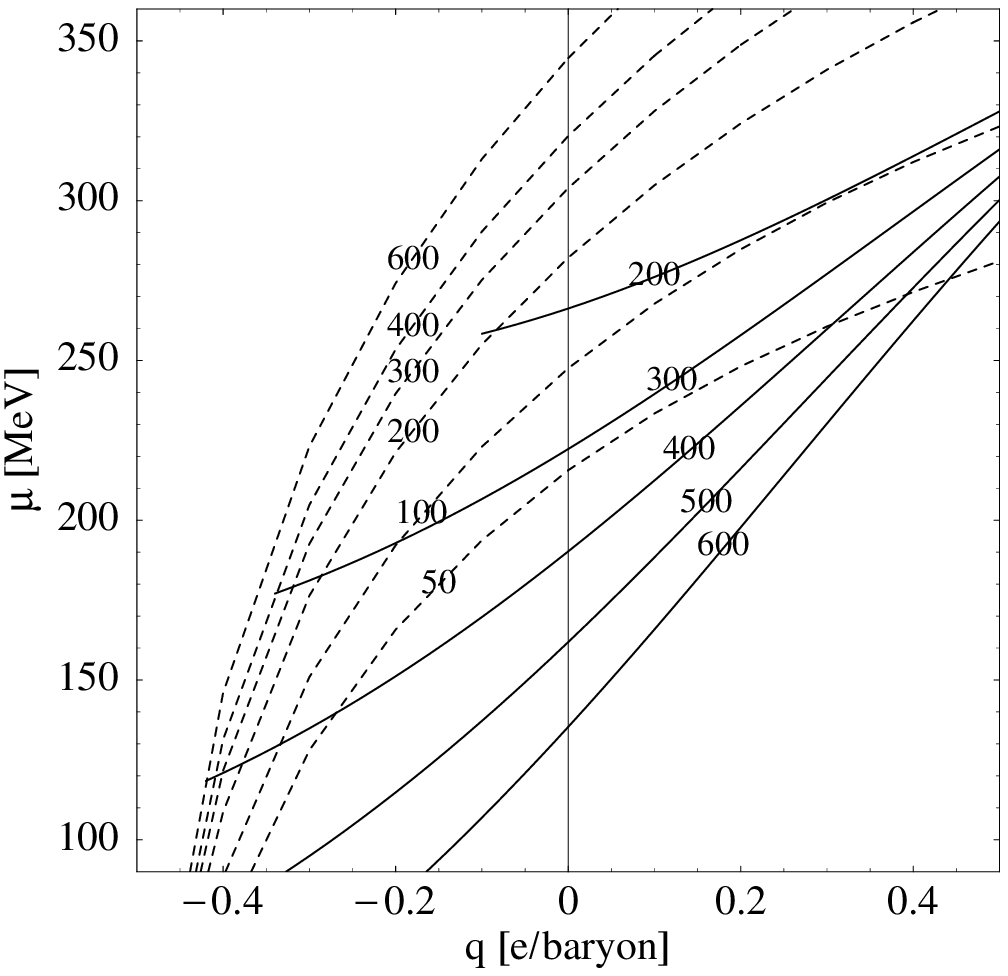}
\includegraphics[width=.43\textwidth]{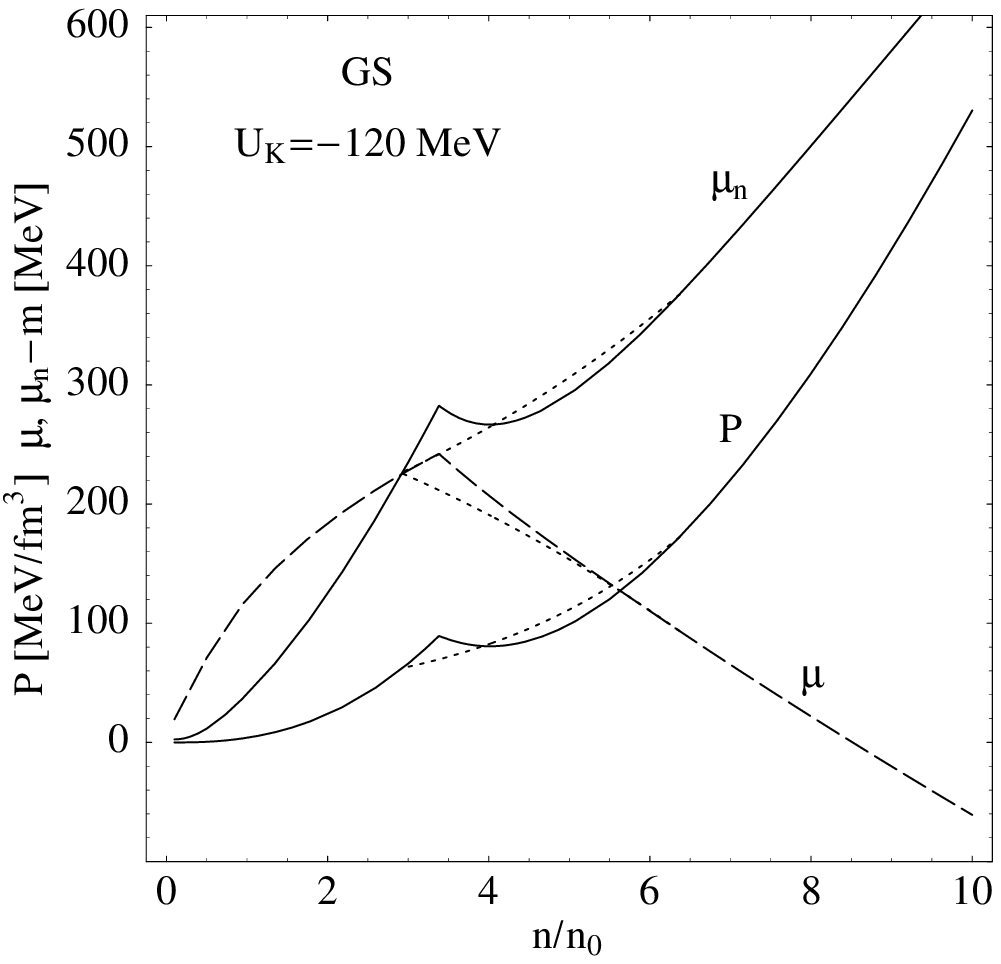}$~~$
\includegraphics[width=.428\textwidth]{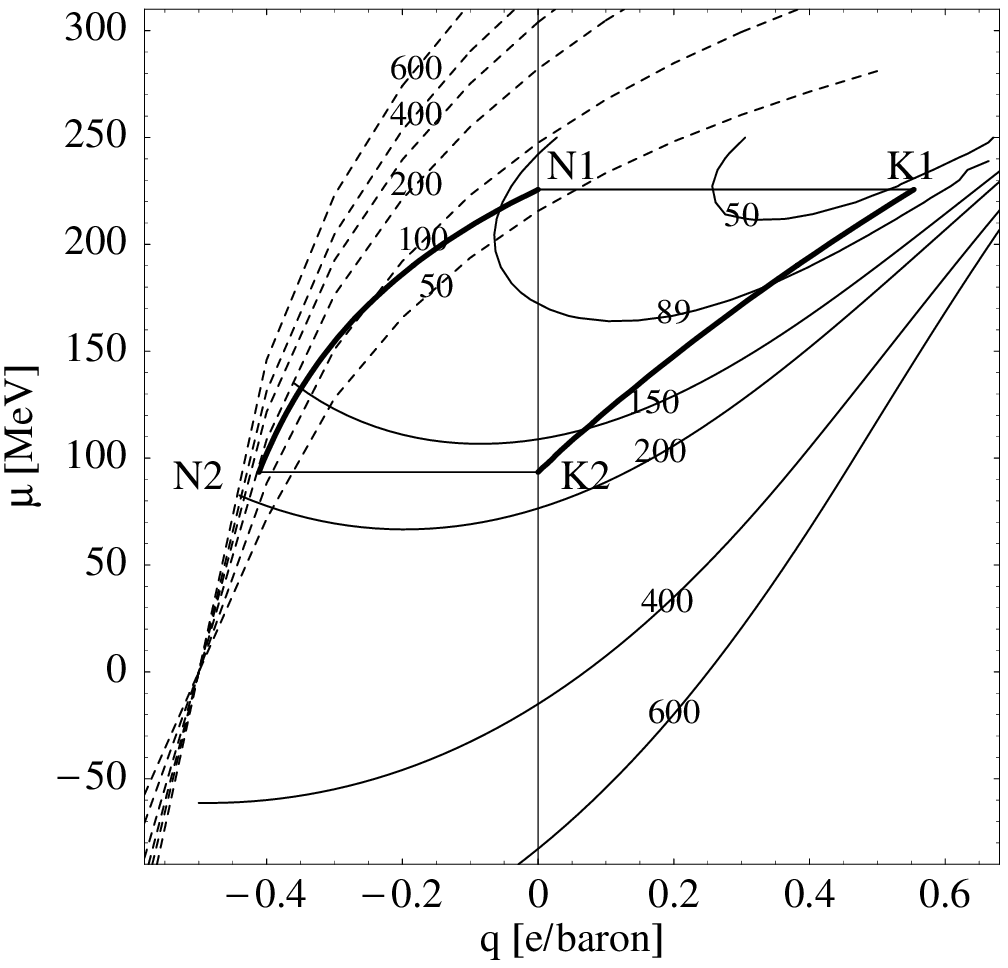}
\caption{Kaon phase stability for the GS model. The left-hand panels show
quantities for neutral phase, the dotted lines representing the values for
the mixed phase region. In the right-hand panels, isobars for the normal 
and kaon phases are plotted. For the $U_K=-120$ case, the phase separation 
is also shown.}
\label{fig-isobGS}
\end{figure*}

\begin{figure*}[!]
\includegraphics[width=.43\textwidth]{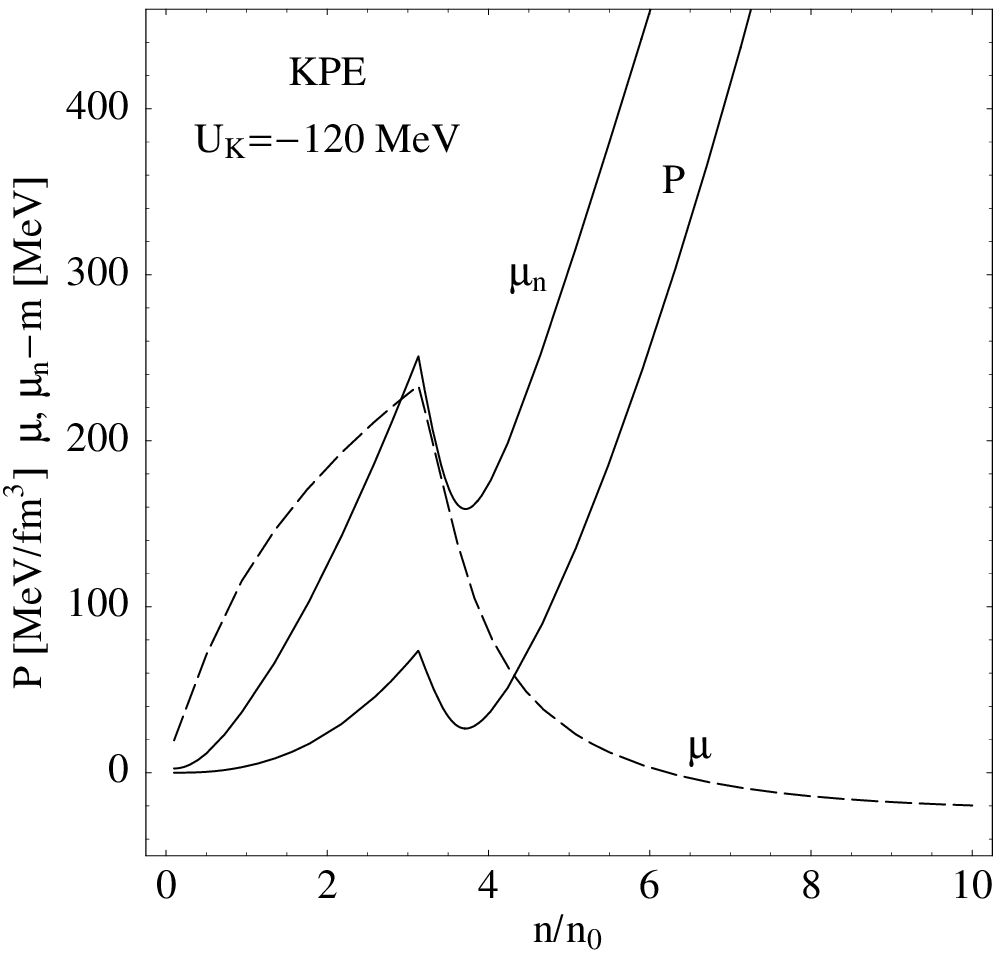}$~$
\includegraphics[width=.422\textwidth]{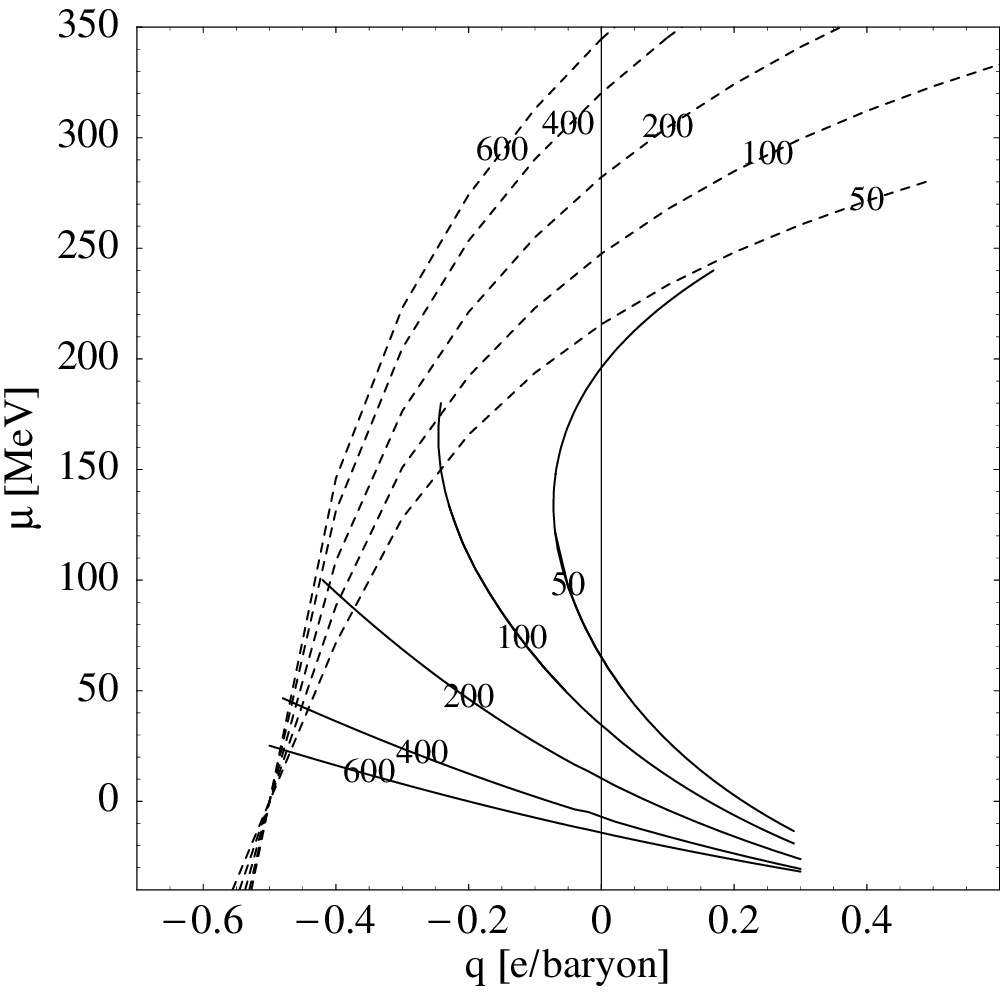}\\
\includegraphics[width=.43\textwidth]{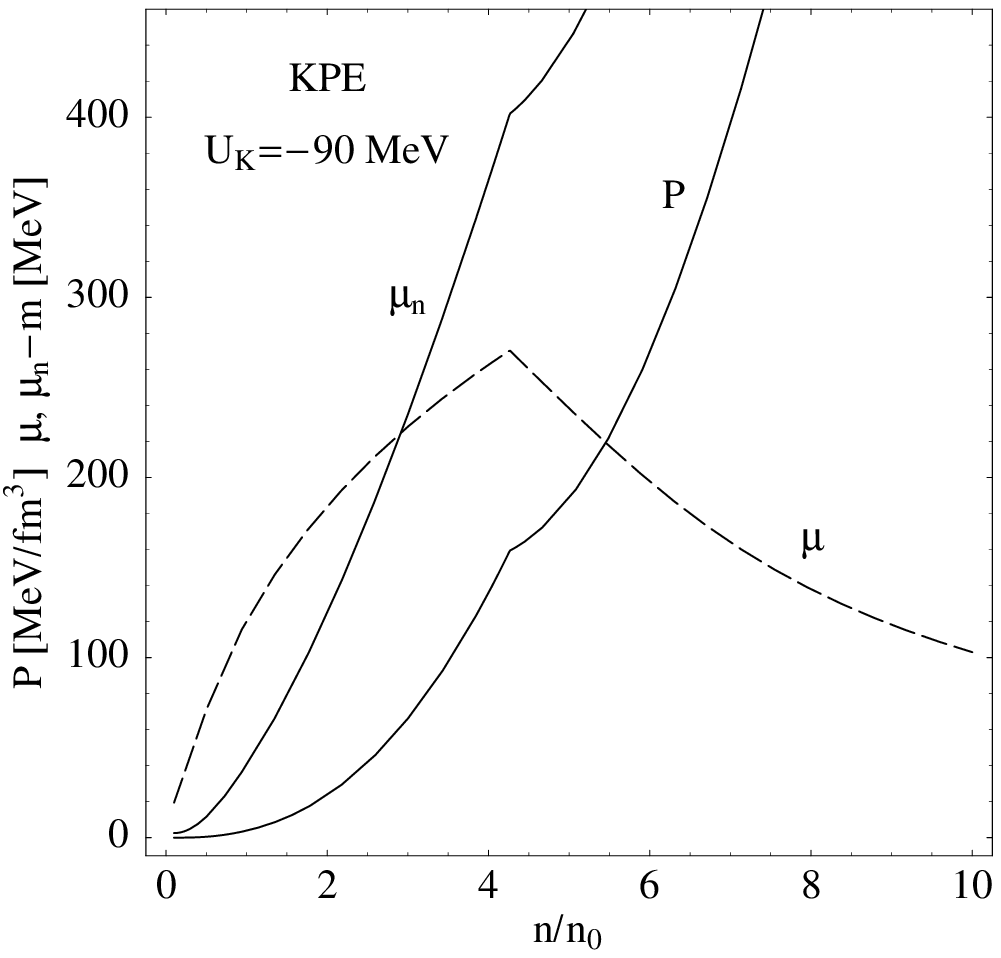}$~~$
\includegraphics[width=.428\textwidth]{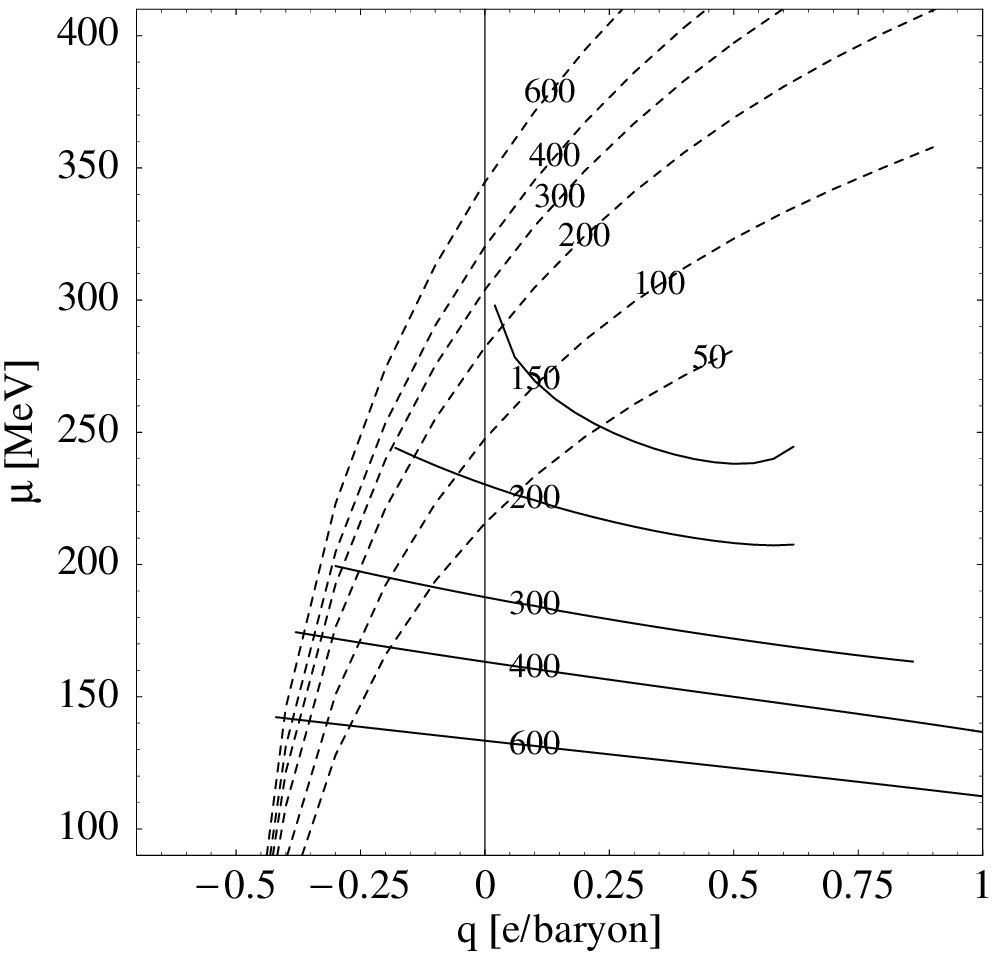}
\caption{The kaon phase instability for the KPE model with optical 
potential $U_K=-120$ and $-90$ MeV.}
\label{fig-isobKPE}
\end{figure*}

The models formulated above are treated in the spirit of RMF theory,
taking into account that the kaon mean field value is $K^{\pm}=\frac{f
\th}{\sqrt{2}} \exp^{\pm i \mu t}$. In order to find the pressure contours
in the $q$-$\mu$ plot, one has to solve the relevant equation of motion
including the beta stable matter (see \cite{Glendenning:1997ak}) for given
$P$ and charge per baryon $q$, where
\beq
q = \frac{-nx +n_{lep} + n_K}{n}
\eeq
and the kaon number density is 
\beq
n_K = 
2(\mu + g_{\omega K} V_0 + g_{\rho K} R_{0,3}) K^- K^+
\eeq
First, we present the stability considerations for the GS model. In 
Fig.\ref{fig-isobGS} one may compare the results for the two values of the 
optical potential.
For the $U_K=-80$ MeV case, when the neutral kaon phase appears its
compressibility is positive at all densities (the left-hand plot). The
kaon phase is also diffusively stable, its isobars having a positive slope
so that that $\chip>0$. For the $U_K=-120$ MeV case, just after the onset
of condensation the compressibility of the neutral phase and $\chip$ are
both negative. To be more explicit, the isobar $P=89~{\rm MeV/fm^3}$,
corresponding the critical density $n_c$ is plotted to show that it keeps 
the negative slope for \mbox{$q=0$}. The diffusive
instability means that the charge fluctuations increase infinitely leading
to phase separation. The charge is unequally deposited into the two
different phases which can coexist only if the conditions of {\em mutual}
stability, expressed by equations (\ref{gibbs_mu}) and (\ref{gibbs_P}),
are fulfilled. The coexistence line (absent in the chiral approach) can be
found for the GS model from the equation
\beq
P^N(\mu_n,\mu)=P^K(\mu_n,\mu)~.
\eeq
In the $q$-$\mu$ plot, the coexistence line corresponds to two lines:  
N1-N2 and K1-K2 because they are located on two different pressure sheets
$P^N(q,\mu)$ and $P^K(q,\mu)$. Proceeding with increasing baryon density,
at the point N1 the system splits up into a nucleon phase and a kaon phase
(the point K1). At the point K1, the kaon phase is maximally charged with
a negative charge ($q$ is positive for negative charge). Increasing the
density further from K1 to K2, the kaon phase loses its charge in favour
of the nucleon phase, becoming neutral at the point K2. At this point the
pure kaon phase starts to be preferred, the nucleon phase vanishes and
again the system is in a one-phase state. We want to point out that the
lines for both the normal and (particularly) the kaon phase pass through a
region where the phases are intrinsically stable, with the corresponding
isobar slopes always being positive, ensuring that $\chip>0$. In the mixed
phase region, the charge $q$ and the baryon number density $n$ are
unequally distributed on the micro-scale. Within a sufficiently large
volume $V$, one may observe droplets of one phase immersed in the other.
The volume fraction $\alpha=V^K/V$ for the two phases is found from the
requirement of the global neutrality:
\beq
\alpha\frac{Q^K}{V^K} + (1-\alpha)\frac{Q^N}{V^N}=
\alpha\,q^K n^K + (1-\alpha)\, q^N n^K =0 .
\eeq
We may then define the average baryon number density for the mixed phase 
region 
\beq
\bar{n} = \alpha\, n^K + (1-\alpha)\, n^N
\eeq
and represent the other thermodynamical quantities ($P, \mu, \ep$ etc.)  
as functions of the one variable $\bar{n}$ (see the dotted lines in the
left-hand panel in Fig.\ref{fig-isobGS}). The above analysis confirms that
for the GS model the intrinsic stability conditions for a given phase
(\ref{positivity1}) and the mutual stability conditions for the mixed
phase (\ref{gibbs_mu},\ref{gibbs_P}) are simultaneously fulfilled. As we
will see in the following, this does not happen for other meson exchange
models like the KPE one. The results for the KPE model are very similar to
those for the chiral model. Results for the two cases $U_K-=-120,-90$ are
shown in Fig.\ref{fig-isobKPE}.
In the first case, negative compressibility appears for neutral phase just
after the critical density. It is impossible to find a region in the
$q$-$\mu$ plane where the kaon phase would be diffusively stable. Even at
higher densities, corresponding to negative values of $\mu$, when the
compressibility of \mbox{$q=0$} phase again becomes positive the kaon phase is
still diffusively unstable. As in the chiral case, this property of the
KPE model remains unchanged also for smaller values of the kaon-nucleon
coupling. As one may see in Fig.\ref{fig-isobKPE} for $U_K=-90$ (we took
this value instead of -80 so that the effect is more visible), in spite of
positive $\kappa_q$ for the whole range of densities considered, the
electrical capacitance $\chip$ is always negative apart from in a small
range of $q$ around 0.6 which corresponds very small values of the
pressure.

\section{Summary and discussion}
In this paper we have focused on the diffusive stability of kaon
condensates within the framework of three different models of strong
interactions: one chiral and two meson-exchange based models (here
referred to as GS and KPE). From this analysis, one can understand why the
Gibbs construction, which can be made in the GS model, is not applicable
to the other models presented here. From a thermodynamical point of view,
the chiral and KPE Lagrangians lead to matter which does not fulfil one
of the fundamental conditions for intrinsic phase stability. This happens
independently of the value of the kaon-nucleon coupling strength and must
be an intrinsic property of the model. At this point it is difficult to
say definitely what are the requirements which should be imposed on a
Lagrangian in order to get complete stability. In the work by Pons et. al
\cite{Pons:2000iy}, the authors considered several meson exchange models
and found that making the Gibbs construction is possible for some of them.
One may suppose that the existence of a mixed phase existence means that
the kaon phase is intrinsically stable, but in principle, this should be
checked independently for the models presented. Looking at the properties
of the Lagrangian for the admissible models, one may suppose that nature
of the kaon to vector-meson coupling is the most relevant property. Making
the Gibbs construction was possible only for models with the covariant
derivative for the kaons given by eq.(\ref{cov-der}), which leads to
vanishing divergence of the vector field. However this condition is not
sufficient; 
as was shown in \cite{Pons:2000iy}, there was also a model with this
property for which the Gibbs construction was not possible.
It seems that the kaon to scalar meson coupling is also
important. To reach a more solid conclusion on this will require further
analysis.
\begin{acknowledgments}
I am pleased to thank J.C.Miller for his warm hospitality at SISSA and 
careful reading of the manuscript, also S.Fantoni and M.Fabbrichesi
for helpful discussions. 
This work was supported by the NATO Fellowship Programme
and the Polish State Committee for Scientific Research Grant 2P03D 02025.
\end{acknowledgments}


\bibliography{stabrev}

\begin{thebibliography}{99}

\bibitem{Kaplan:yq}
D.~B.~Kaplan and A.~E.~Nelson,
Phys.\ Lett.\ B {\bf 175} (1986) 57.


\bibitem{Lee:ef}
C.~H.~Lee,
Phys.\ Rept.\  {\bf 275} (1996) 255.

\bibitem{CCWZ}
S.~R.~Coleman, J.~Wess and B.~Zumino,
Phys.\ Rev.\  {\bf 177} (1969) 2239.\\
C.~G.~Callan, S.~R.~Coleman, J.~Wess and B.~Zumino,
Phys.\ Rev.\  {\bf 177} (1969) 2247.


\bibitem{Mueller-Groeling:cw}
A.~Mueller- Groeling, K.~Holinde and J.~Speth,
Nucl.\ Phys.\ A {\bf 513} (1990) 557.

\bibitem{Knorren:1995ds}
R.~Knorren, M.~Prakash and P.~J.~Ellis,
Phys.\ Rev.\ C {\bf 52} (1995) 3470

\bibitem{Kubis:2002dr}
S.~Kubis and M.~Kutschera,
Nucl.\ Phys.\ A {\bf 720} (2003) 189

\bibitem{Glendenning:1992vb}
N.~K.~Glendenning,
Phys.\ Rev.\ D {\bf 46} (1992) 1274.

\bibitem{callen}
Herbert B. Callen {\em Thermodynamics}, John Wiley \& Sons, Inc.,
New York (1960)

\bibitem{Glendenning:1997ak}
N.~K.~Glendenning and J.~Schaffner-Bielich,
Phys.\ Rev.\ C {\bf 60} (1999) 025803


\bibitem{Thorsson:bu}
V.~Thorsson, M.~Prakash and J.~M.~Lattimer,
Nucl.\ Phys.\ A {\bf 572} (1994) 693
[Erratum-ibid.\ A {\bf 574} (1994) 851]

\bibitem{Donoghue:1985bu}
J.~F.~Donoghue and C.~R.~Nappi,
Phys.\ Lett.\ B {\bf 168} (1986) 105.

\bibitem{Lattimer61}
M.~Prakash, T.~L.~Ainsworth and  J.~M.~Lattimer,
Phys.\ Rev.\ Lett.\  {\bf 61} (1988) 2518.

\bibitem{Pons:2000iy}
J.~A.~Pons, S.~Reddy, P.~J.~Ellis, M.~Prakash and J.~M.~Lattimer,
Phys.\ Rev.\ C {\bf 62} (2000) 035803

\bibitem{Schaffner:1995th}
J.~Schaffner and I.~N.~Mishustin,
Phys.\ Rev.\ C {\bf 53} (1996) 1416

\bibitem{Kubis:1997ew}
S.~Kubis and M.~Kutschera,
Phys.\ Lett.\ B {\bf 399} (1997) 191

\end{thebibliography}

\end{document}